# Electrical transport properties of atomically thin $WSe_2$ using perpendicular magnetic anisotropy metal contacts

S. Gupta[1,*] R. Ohshima[1], Y. Ando[1], T. Endo[2], Y. Miyata[2], and M. Shiraishi[1]

[1]Department of Electronic Science and Engineering, Kyoto University, Kyoto 615-8510, Japan

[2]Department of Physics, Tokyo Metropolitan University, Tokyo 192-0397, Japan

[$]Corresponding author: Sachin Gupta (sachin.gupta@bennett.edu.in)

Tungsten diselenide, $WSe_2$ shows excellent properties and become very promising material among two dimensional semiconductors. Wide band gap and large spin-orbit coupling along with naturally lacking inversion symmetry in the monolayer $WSe_2$ make it efficient material for spintronics, optoelectronics and valleytronics applications. In this work, we report electrical transport properties of monolayer $WSe_2$ based field effect transistor with most needed multilayer Co/Pt ferromagnetic electrodes exhibiting perpendicular magnetic anisotropy. We studied contacts behaviour by performing *I-V* curve measurements and estimating Schottky barrier heights (SBHs). SBHs estimated from experimental data are found to be comparatively small, without using any tunnel barrier. This work expands the current understanding of $WSe_2$ based devices and gives insight into the electrical behaviour of Co/Pt metal contacts, which can open great possibilities for spintronic/valleytronic applications.

Scaling issues and short channel effects in silicon-based electronics and exfoliation of graphene in atomically thin layer form has led the quest of two-dimensional (2D) semiconducting materials[1,2]. Transition metal dichalcogenides (TMDs) is a large family of 2D layered materials, which show various physical properties ranging from semiconducting to superconducting[3]. Among these materials, semiconducting materials such as $MoS_2$ and $WSe_2$ has have drawn considerable interest due to their direct band gap semiconducting nature in monolayer form associated with attracting electronic, mechanical, optical, and valleytronic properties and therefore proposed for various technological applications[2,3,4]. These materials are among the best candidates for valleytronic applications because in monolayer form they naturally lack inversion symmetry and enable easy dynamical control of valley degree of freedom. Although all of them show a significant amount of spin-orbit splitting in their valence bands, however splitting for $WSe_2$ is larger compared to other TMDs due to possession of heavier atoms[5]. Bands with large spin splitting in $WSe_2$ can make it a promising candidate to study valley dependent spin properties in spintronics/valleytronics.

Due to semiconducting nature of $WSe_2$ and formation of metal-semiconductor junction during device fabrication, these devices show large Schottky barriers[6]. Large Schottky barrier height (SBH) impedes injection of charge carriers and could make it difficult to observe valley dependent spin signals. In order to understand/improve electrical properties of $WSe_2$ based field effect transistor (FET), researchers have used metal electrodes with different work functions, intentionally doped $WSe_2$ channel or electrodes, used 2D electrodes and tunnel barriers[7,8,9,10]. Since spin polarization in $WSe_2$ is perpendicular to the sample plane, perpendicular magnetic anisotropy (PMA) electrodes are the pre-requisite to detect spin-valley signals and integrate $WSe_2$ material in valleytronic/spintronic devices. There are very limited reports in the literature, which discuss transport properties of monolayer $WSe_2$ FET using ferromagnetic electrodes (especially PMA electrodes)[9]. As transport properties of $WSe_2$ materials depend on growth method, quality of interface, metal work function and other fabrication procedure, it is very important to explore these properties more with different metal electrodes. In this paper, we fabricated FET using salt-assisted CVD grown single layer $WSe_2$ channels and multilayer Co/Pt metal electrodes, which show PMA character. We studied electrical properties of back-gated monolayer $WSe_2$ at different temperatures.

Atomically thin $WSe_2$ material was grown via salt-assisted chemical vapor deposition (CVD) technique on *n*-type doped Si/$SiO_2$ substrate, where thickness of thermally oxidized $SiO_2$ was ~285 nm. TMDs grown using this method was found to show excellent physical properties[11,12,13,14]. To create $WSe_2$ channels (6x4 $\mu m^2$), we spin-coated TGMR resist (negative tone) and performed electron-beam (EB) lithography. Protecting $WSe_2$ channel by TGMR, we etched unwanted $WSe_2$ from substrate using $O_2$ plasma etching. Later, TGMR was lifted by wet etching using N-Methyl-2-pyrrolidone and rinsed by acetone and isopropyl alcohol (IPA). Once channel is formed, we deposited Co/Pt multilayer (PMA) electrodes via magnetron sputtering at room temperature[15]. We confirmed PMA of electrodes on $SiO_2$ substrate[15] and on $WSe_2$ sample by the Hall measurement with a Hall bar structure. Ti(3nm)/Au(60nm) was evaporated via EB deposition technique to form pads in our FET devices. Confirmation of single layer $WSe_2$ channels was done by Raman spectroscopy experiment using a laser light of wavelength of 488 nm. Multi-terminal electrical measurements were carried out using helium-free cryostat.

A Hall bar structure was patterned using EB lithography on single layer $WSe_2$ grown on $SiO_2$/Si substrate to confirm PMA. Multilayer $[Co(0.5)/Pt(3.4)]_2$ (numbers in parentheses are the layer thickness in nm) structure was sputtered at room temperature and Hall measurement was carried out using physical property measurement system (Quantum Design, USA). Hall bar and measurement set-up is shown in Fig. 1(a). Fig. 1(b) shows perpendicular field ($H_\perp$) dependence of Hall resistance $R_{xy}$ at 5 K. The plot shows sharp magnetic transition with field sweep in up and down direction with clear hysteresis, which confirms PMA of multilayer Co/Pt superlattice structure.

Fig. 2 shows four-probe *I-V* curve measurements performed at room temperature to calculate channel resistance of monolayer $WSe_2$. A bias voltage of 2V was applied at the outer electrodes, current flowing through the outer electrodes was recoded as a function of voltage drop between two inner electrodes. Four-probe *I-V* curve shows linear behaviour in the entire range of applied bias voltage. The channel resistance estimated from results in Fig. 2 is of the order of several hundred killoOhm. High channel resistance in $WSe_2$ is the indicative of large bandgap in these materials.

We carried out two-probe drain-source current – drain-source voltage (($I_{DS}$ - $V_{DS}$) measurements at room temperature as a function of back gate voltage ($V_g$). The back gate voltage was applied by scratching back side of Si substrate. Maximum 10 V of back gate voltage was applied to keep leakage current less than 1 nA. Obtained results are plotted in Fig. 3. $I_{DS}$ - $V_{DS}$ curves exhibit symmetric behaviour as expected due to back-to-back diode like structure formed at source and drain of the FET. One can note from Fig. 3(a) that $I_{DS}$ - $V_{DS}$ curves show deviation from linearity, which suggests formation of Schottky barriers at metal-semiconductor junction. It is worth to note that $I_{DS}$ - $V_{DS}$ curves are almost same even after applying back gate voltage from -10 V to 10 V, which reflects that back gate is not sufficient to modulate barriers at the interface. Fig. 3(b) shows back gate voltage dependence of $I_{DS}$ at different values of bias voltages. Very weak dependence of $I_{DS}$ upon back gate voltage can be clearly seen. The observation in the present case is similar to Ti contacts reported previously[16]. Negligible increase in $I_{DS}$ with back gate voltage reflects possible Fermi level pinning. Strong electrostatic gating is required to make it clear whether device is *n* type, *p* type or showing ambipolar behaviour. Usually, monolayer $WSe_2$ have been reported to show ambipolar behaviour, however in the case of heavy metal electrodes such as Pd, it was found to exhibit *p*-type behaviour as heavy metals have large work function, which can inject hole carriers in valence band of $WSe_2$[16]. For low work function metal contacts such as Al, $WSe_2$ was reported to show *n*-type device characteristics[17].

To shed light on the Schottky barrier formed at the metal semiconductor interface, we performed $I_{DS}$ - $V_{DS}$ curves measurements as a function of temperature in our back-gated device. SBH was calculated by employing thermionic emission model modified for 2D materials[18] give by

$$I_{DS} = AA^{*}T^{\frac{3}{2}}\exp\left[-\frac{e}{k_B T}\left(\phi_B - \frac{V_{DS}}{n}\right)\right] , \qquad (1)$$

where $A$, $A^*$, $k_B$ are the contact surface area, 2D equivalent Richardson constant and Boltzmann constant, respectively. Here ideality factor is denoted by $n$ and Schottky barrier height (in eV) by $e\phi_B$. Equation (1) can be re-arranged in the form given below

$$ln\left(I_{DS}/\,T^{\frac{3}{2}}\right) = \ln A + lnA^{*} - \frac{E_A}{k_B}\left(\frac{1}{T}\right). \qquad (2)$$

Where $E_A$ is an activation energy and is given as $E_A = e(\phi_B - \frac{V_{DS}}{n})$. To contribute current in the FET device, charge carriers should overcome activation energy to cross the barrier and move through the $WSe_2$ channel.

SBH can be estimated experimentally by Arrhenius plot, $ln\left(I_{DS}/\,T^{\frac{3}{2}}\right)$ vs. $1/T$ as shown in Fig. 4(a). The slope of plot was estimated by fitting equation (2) to the experimental data for various values of $V_{DS}$ at $V_g = 0$ V. The slope is given by $-E_A/1000k_B$ and is plotted as a function of $V_{DS}$ in Fig. 4 (b) and (c) for $V_g = 0$ and 10V, respectively. The intercepts from the linear fits to the experimental data gives SBH. The error for SBH at 10 V is larger due to scattering points. The value of SBH was found to be 36.7 and 38.3 meV for $V_g = 0$ and 10V, respectively. The observed value of SBH is smaller than various metal contacts used before to fabricate $WSe_2$ based FET. The previously reported SBH values for Pt[9] and Pd[6] contacts are ~239 meV and Pd ~325 meV, respectively. We used multilayer [Co/Pt] metal contacts. It is anticipated that when contacts are multilayer structures, the work function at the interface is modified and is given by the effective work function[19]. This can affect the SBH at the interface. It is worth to note that the value of SBH is almost same at 0 and 10 V, there is hardly any modulation using gate, which was also reflected in *I-V* curves in Fig. 3.

Electrostatic gating can modulate Schottky barriers formed at the metal-semiconductor interface either by band bending and/or by electrical doping into the semiconductor channel[15,20]. According to Schottky- Mott model, SBH in a metal-semiconductor junction linearly depends on metal work function, however it has been observed that in most of cases, SBH is insensitive to the metal work function and Fermi level is pinned in the gap of semiconductor which is less tuneable with respect to different work function of the metals used as an electrodes[21]. The Fermi level pinning (FLP) in the system can be either caused by formation of metal induced gap states in the semiconductor or interface dipole or defects at the interface[21] and can affect the SBH dramatically. Strong FLP in the system might be a reason of SBH not responding to the electrical gating. To calculate interfacial states and know the exact reason, we require further experiments with more experimental data with various metal contacts and models, however, is out of scope of this paper.

Multilayer Co/Pt structure was deposited to develop ferromagnetic electrodes with perpendicular magnetic anisotropy (PMA) and integrated to fabricate monolayer $WSe_2$ based field effect transistor. We studied electrical transport properties using two and four probe measurement configurations. Two- probe *I-V* curves show very weak gate dependence and suggest ambipolar behaviour in our FET device. We also estimated Schottky barrier height (SBH) by performing temperature dependence of *I-V* curves of back-gated FET. SBH was found to be independent of gate voltage. The present study can help scientists in further research in $WSe_2$ based devices using PMA electrodes, which is very important for emerging application of valleytronics.

The authors thank Prof. T. Kimoto and K. Kanegae for allowing us to access Raman spectroscopy facility. This work was supported by a Grant-in-Aid for Scientific Research (S) No. 16H06330, “Semiconductor spincurrentronics”, and MEXT (Innovative Area “Nano Spin Conversion Science” KAKENHI No. 26103003). Y. M acknowledges the financial support from JST CREST (Grant No. JPMJCR16F3).

## Figure captions:

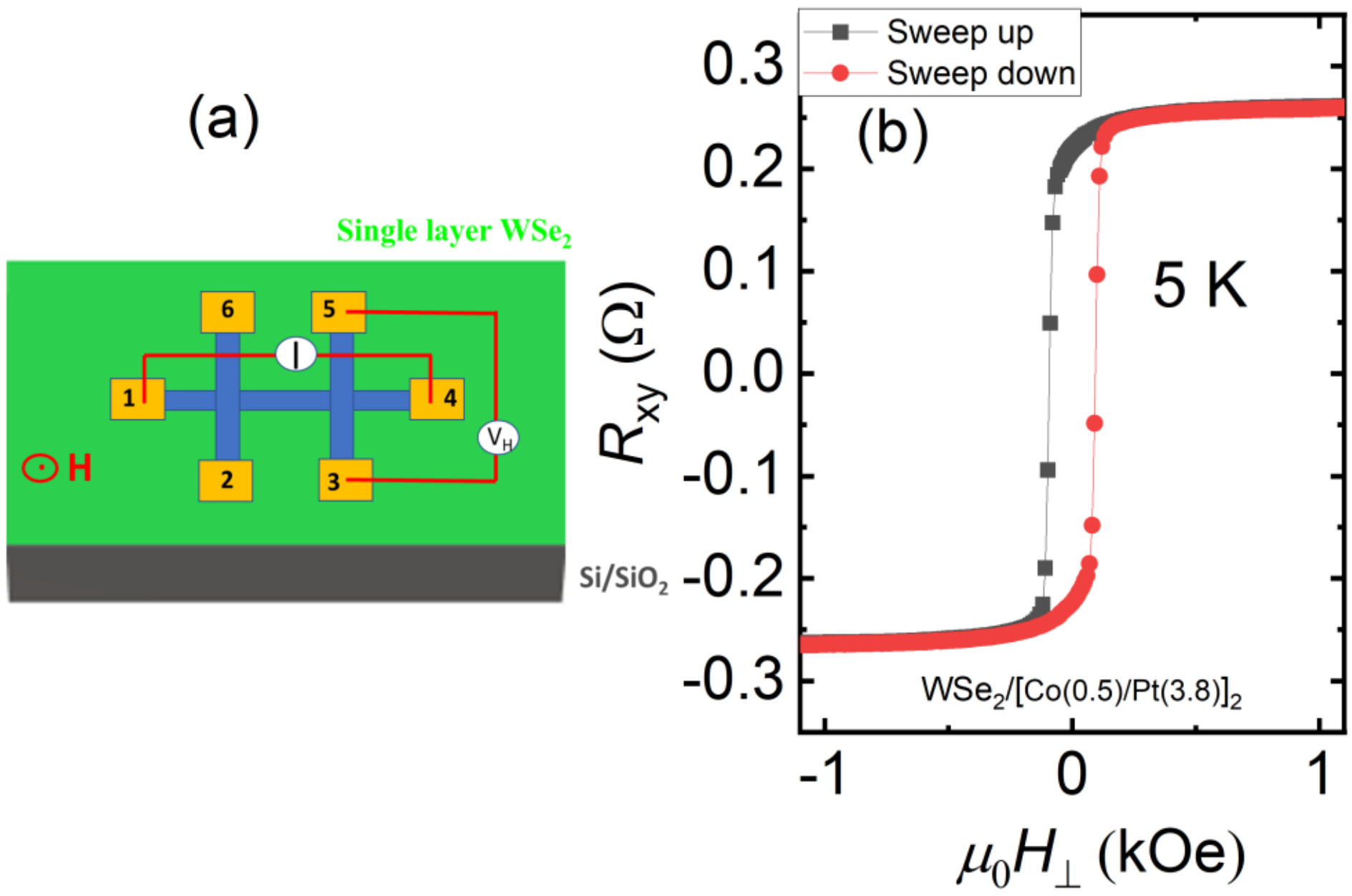

**Fig. 1. (a)** Hall bar pattern of PMA on single layer $WSe_2$. The current is applied between contacts 1and 4 and Hall voltage, $V_H$ is measured between contacts 3 and 5. Magnetic field is applied perpendicular to sample plane. (b) Field dependence of Hall resistance, $R_{xy}$ of a Hall bar device at 5 K.

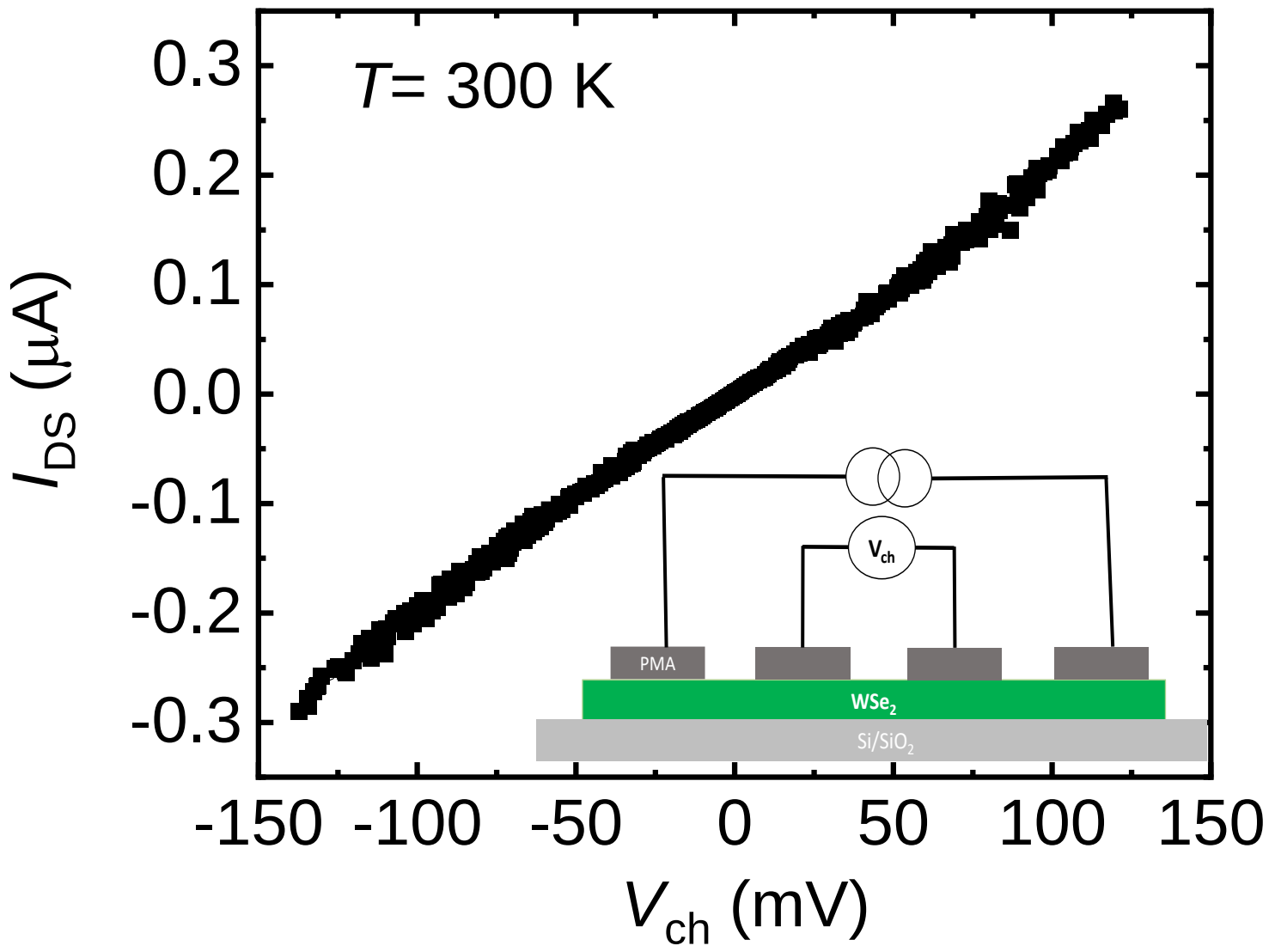

**Fig. 2.** Four-probe *I-V* characteristics of monolayer $WSe_2$ FET at room temperature. The inset shows measurement set-up in which current is applied on the outer electrodes and voltage drop (defined as channel voltage, $V_{ch}$) is measured between inner electrodes.

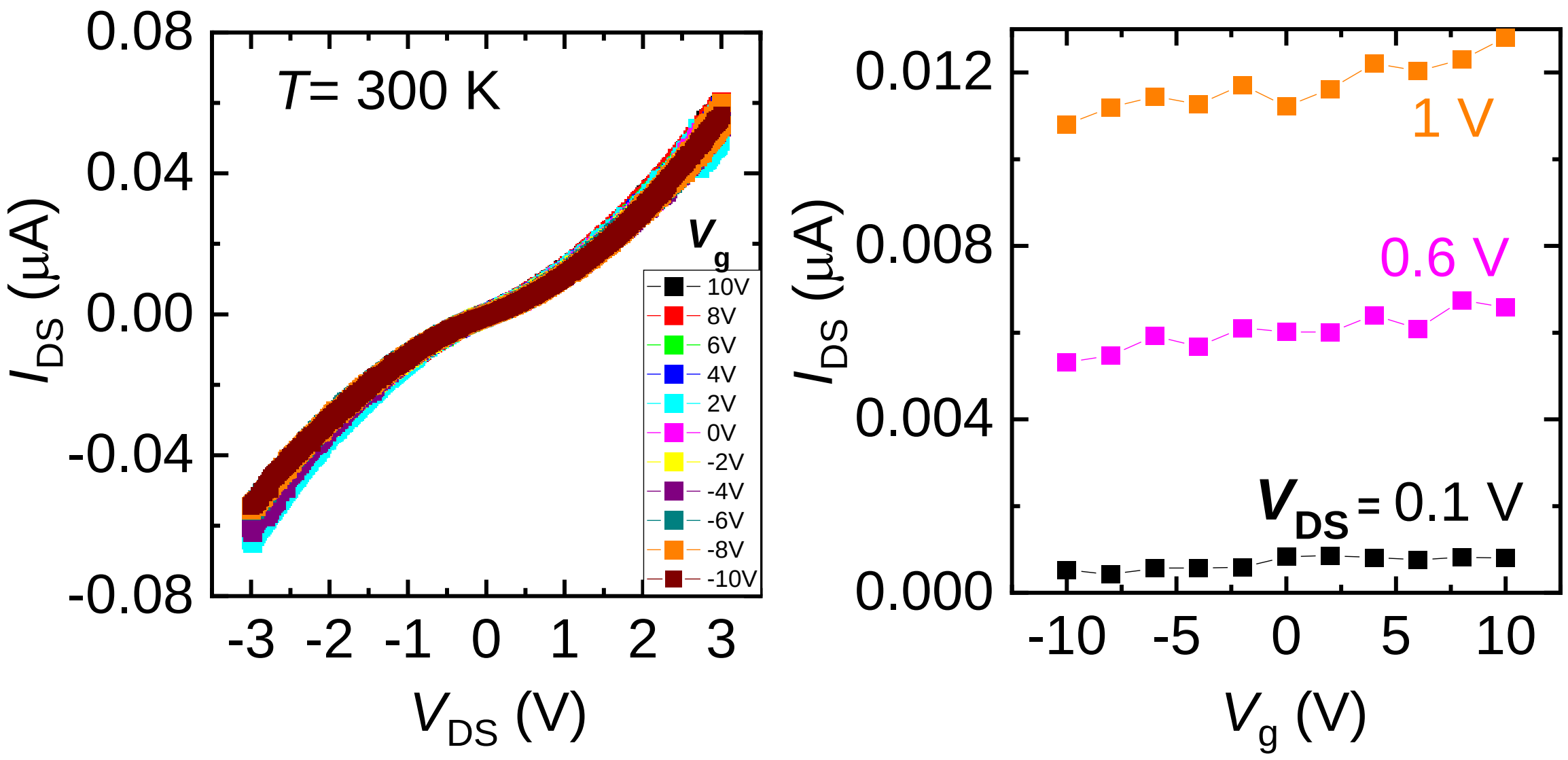

**Fig. 3.** (a) $I_{DS}$ vs $V_{DS}$ curves as a function of back gate voltage ($V_g$) recorded at room temperature. (b) $V_g$ dependence of $I_{DS}$ at different bias voltages ($V_{DS}$).

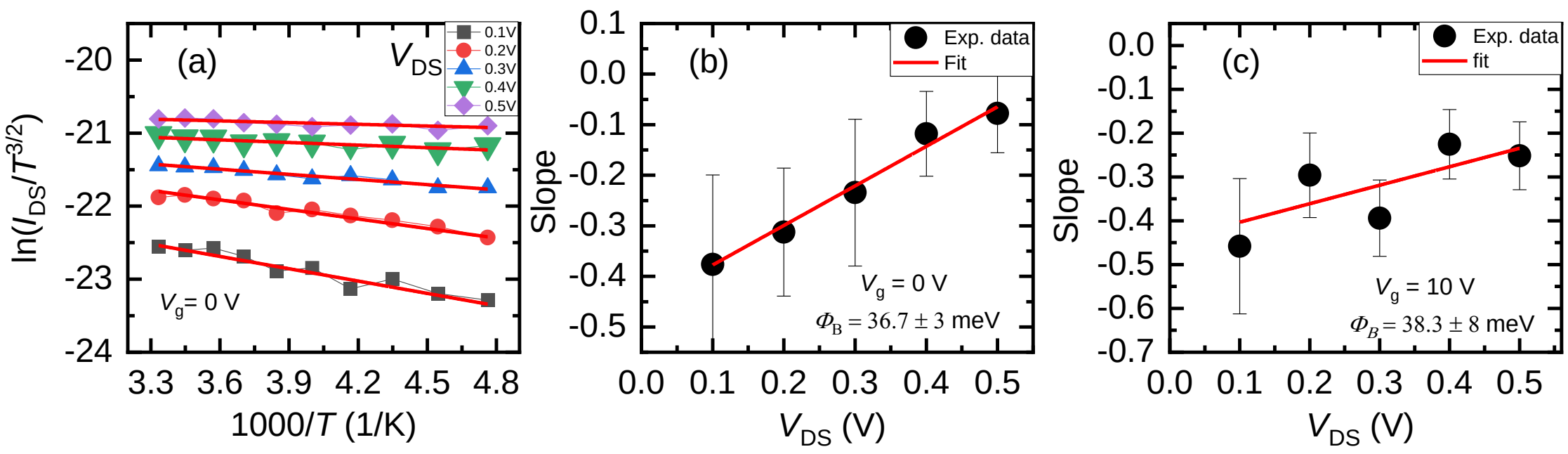

**Fig. 4.** (a) The Arrhenius plot (ln $I_{DS}/T^{3/2}$ vs. $1000/T$) for various bias voltages ($V_{DS}$) at. $V_g$ = 0V. (b) Schottky barrier heights (SBH) calculated at (b) $V_g$ = 0V and (c) $V_g$ = 10V.